\documentclass[epj]{svjour}
\usepackage{graphics}
\begin{document}
\titlerunning{Electrical Conduction of Ti/TiO$_{x}$/Ti Structures}
\title{Electrical Conduction of Ti/TiO$_{x}$/Ti Structures at Low Temperatures and High Magnetic Fields}
%\subtitle{Do you have a subtitle?\\ If so, write it here}
\author{  Marianna Batkova  and Ivan Batko%\inst{1}% etc
% \thanks is optional - remove next line if not needed
}                     % Do not remove
\institute{Institute of Experimental  Physics,
 Slovak   Academy  of Sciences, Watsonova 47,
 040~01~Ko\v {s}ice, Slovakia}%
\date{Received: date / Revised version: date}
% The correct dates will be entered by Springer
%
\abstract
{We present results of electrical conduction studies of Ti/TiO$_{x}$/Ti planar structures prepared by tip-induced local anodic oxidation of titanium thin films. 
	The prepared structures have shown almost linear $I-V$ curves at temperatures between 300~K and 30~K, and only slight deviation from linear behaviour at lower temperatures.
	Electrical conductance of the structures can be adequately explained by a two-channel model where variable range hopping channels and metallic ones coexist in parallel, while a crossover from Mott to Efros-Shklovskii variable-range-hopping
  conductivity has been observed at decreasing temperature. 
		The magnetoresistance of the studied structures is very small  even in magnetic fields up to 9~T.
		The reported electrical properties of the structures indicate their promising applications 
as very low heat capacity temperature sensors for cryogenic region and high magnetic fields.
} %end of abstract
\maketitle
\section{Introduction}
\label{Intr}

			Thin films and thin-films structures composed of metal oxides display a rich variety of electronic and magnetic properties, including colossal magnetoresistance, superconductivity, and multiferroic behaviour \cite{Mannhart2010,Zegenhagen2015}.	
			Due to diversity of their properties, many metal-oxide based materials and structures find utilization in various technological applications, such as resistance switches \cite{Pham2013}, electrodes \cite{Sabet2015,Jafari2015}, or sensors of temperature \cite{Batko95Cryogenic}, magnetic field \cite{Zutic2004}, or gas \cite{Patil2011}. 
			For example, titanium forms a wide range of stoichiometric oxides with different electrical properties; there can be found metals: TiO, Ti$_{2}$O, as well as semiconductors: Ti$_{6}$O$_{11}$, Ti$_{8}$O$_{15}$, or Ti$_{5}$O$_{9}$, and also compounds showing semiconductor - metal transition: TiO$_{2}$, Ti$_{2}$O$_{3}$, Ti$_{3}$O$_{5}$, or Ti$_{4}$O$_{7}$ \cite{Morin1959,Rao1974}. 
		 Moreover, electrical conductivity of transition metal oxides is markedly affected by deviations from stoichiometry \cite{Rao1974} and therefore, electrical properties of these oxides are strongly influenced by preparation method and parameters of synthesis process.
		
		An interesting alternative to prepare patterns of metal-oxide materials is local anodic oxidation (LAO) of metallic surfaces by use of atomic force microscope (AFM) \cite{Irmer97,Dubois1999,Vullers1999,Davis2003}. 
		 Direct oxidation of a sample by negative voltage applied to an AFM tip, with respect to the oxidized sample, is a  base for LAO,  while the process utilizes presence of a water bridge created between the sample and the tip \cite{Cambel2007,Gomez2003}. 
		For local electric fields larger than critical one, 10$^9$~V/m, the water molecules are dissolved to H$^+$ and OH$^-$ ions.
		Consequently, OH$^-$ ions are transported to the positively biased sample surface (anode) in the direction of the electric field, approach the sample surface, react with the surface atoms, and form oxide objects \cite{Cambel2007}. 
		Of course, properties of oxides synthetized this way are affected by parameters of the LAO process (e.g. bias voltage, AFM mode, speed of AFM-tip movement).
		Therefore LAO enables to prepare structures containing various oxides with different properties. 
		In addition, because of the fact that oxidation process propagates inward from the surface, formation of systems with different content of oxide as a function of depth is possible. 

		In this work we report results of studies of Ti/TiO$_{x}$/Ti test structures prepared by LAO on 15~nm thick titanium films.
		Because of relatively higher thickness of the films, not negligible formation of lower oxides of titanium (with metalic behaviour) is expected besides higher (semiconducting) ones.
 		Therefore, investigated Ti/TiO$_{x}$/Ti structures are believed to be electrically heterogenous, containing metallic, as well as semiconducting phases. 
		Main purpose of this paper is to  describe electrical properties of these composite systems, and indicate possible application fields.

\section{Experiment}
\label{Experiment}
\begin{figure}[!tb]%
				\begin{center}
											\resizebox{1\columnwidth}{!}{%
  			\includegraphics{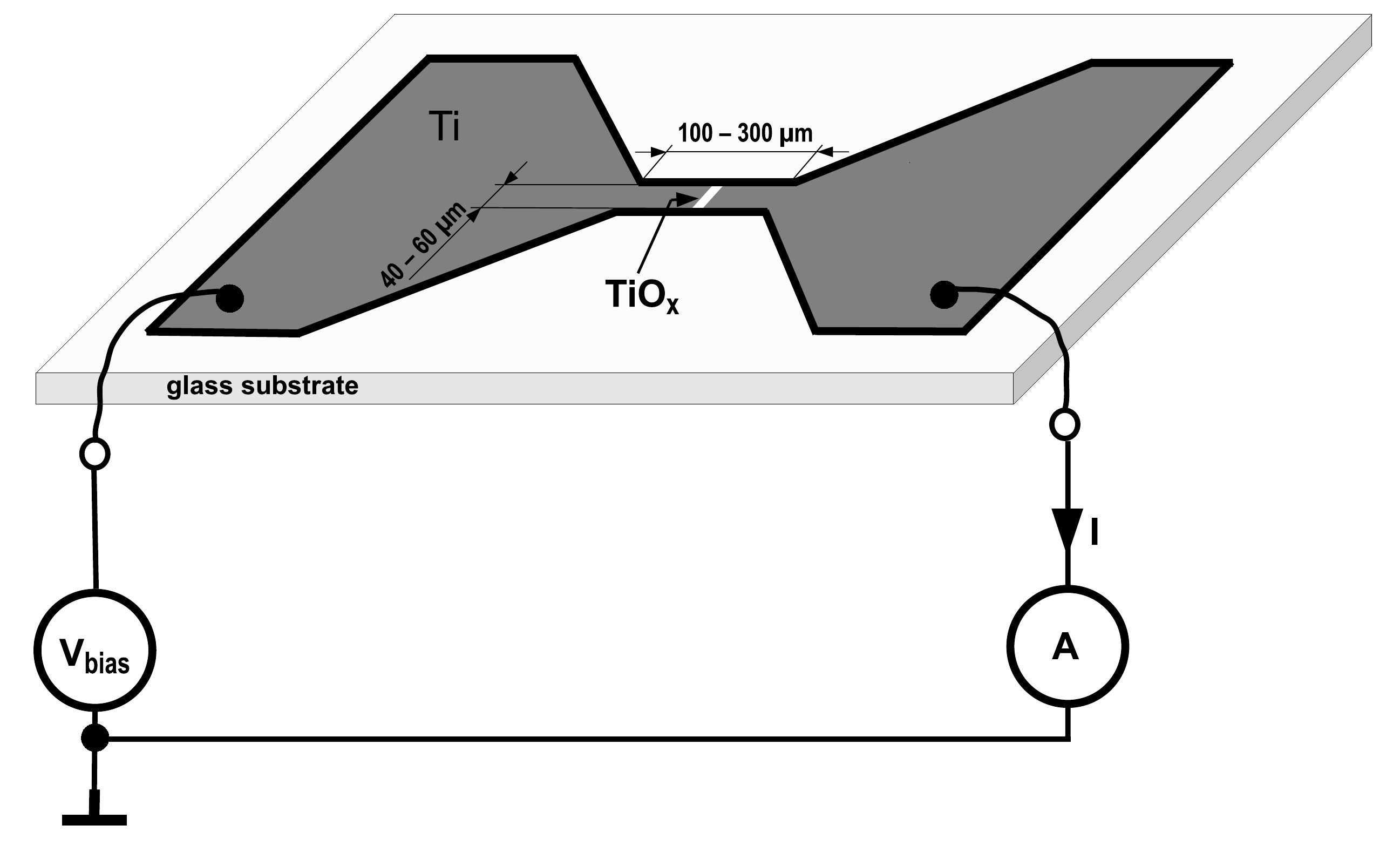}
            }
        \end{center}
\caption{ Schematic depiction of Ti/TiO$_{x}$/Ti structures prepared by tip-induced oxidation of titanium thin film (dimensions are not proportional), and schematic diagram of the circuit used to measure $I-V$ curves. 
	 }		
	 \label{Fig1}
			\end{figure}

			Ti/TiO$_{x}$/Ti test structures, like schematically
depicted in Fig.~\ref{Fig1} were prepared by tip-induced oxidation of titanium thin films as follows.  
			First, microbridges (typically 15~nm thick, $40-60$~$\mu$m wide, and $100-300$~$\mu$m long) were deposited by DC magnetron sputtering on glass substrates kept at ambient temperature. 
			 A shadow mask was used to define geometry of the bridges. 
			 The sputtering was done from polycrystalline Ti target at Ar pressure of 3~mTorr and rate of 0.04~nm/s. 
			 Oxide barriers across the bridges were fabricated by tip-induced LAO using an AFM equipped with a commercial nanolithography software. 
			 The oxidation process was performed in contact mode at ambient conditions, with relative humidity temporarily increased  to a level of $55-60$~$\%$, to ensure a stable tip-induced oxidation process for bias voltage less than 10~V (an instrument limitation). 
			 Only "weak" oxidation was applied, resulting in  resistance increase of the structure due to the oxidation not greater than one order of magnitude. 
			 This is an essential difference in comparison to similar structures reported elsewhere, whith oxidation process repeated until electrical resistance increased by three orders of magnitude \cite{Irmer97,Batko2012EPJ,Soltys2001}. 
			 (Note that such "intensively" oxidized structures reveale non-linear $I-V$ curves \cite{Irmer97,Batko2012EPJ,Soltys2001} and can exhibit memristive behaviour \cite{Batko2012EPJ}.) 
			Measurements of $I-V$ curves were done by applying bias voltage to the structure and measuring corresponding current flowing through the structure, as it is depicted in Fig.~\ref{Fig1}.

\section{Results and discussion}

%%%%%%%%%%%%%%%%%%%%%%%%%%
\begin{figure}[!tb]%
				\begin{center}
											\resizebox{1\columnwidth}{!}{%
  			\includegraphics{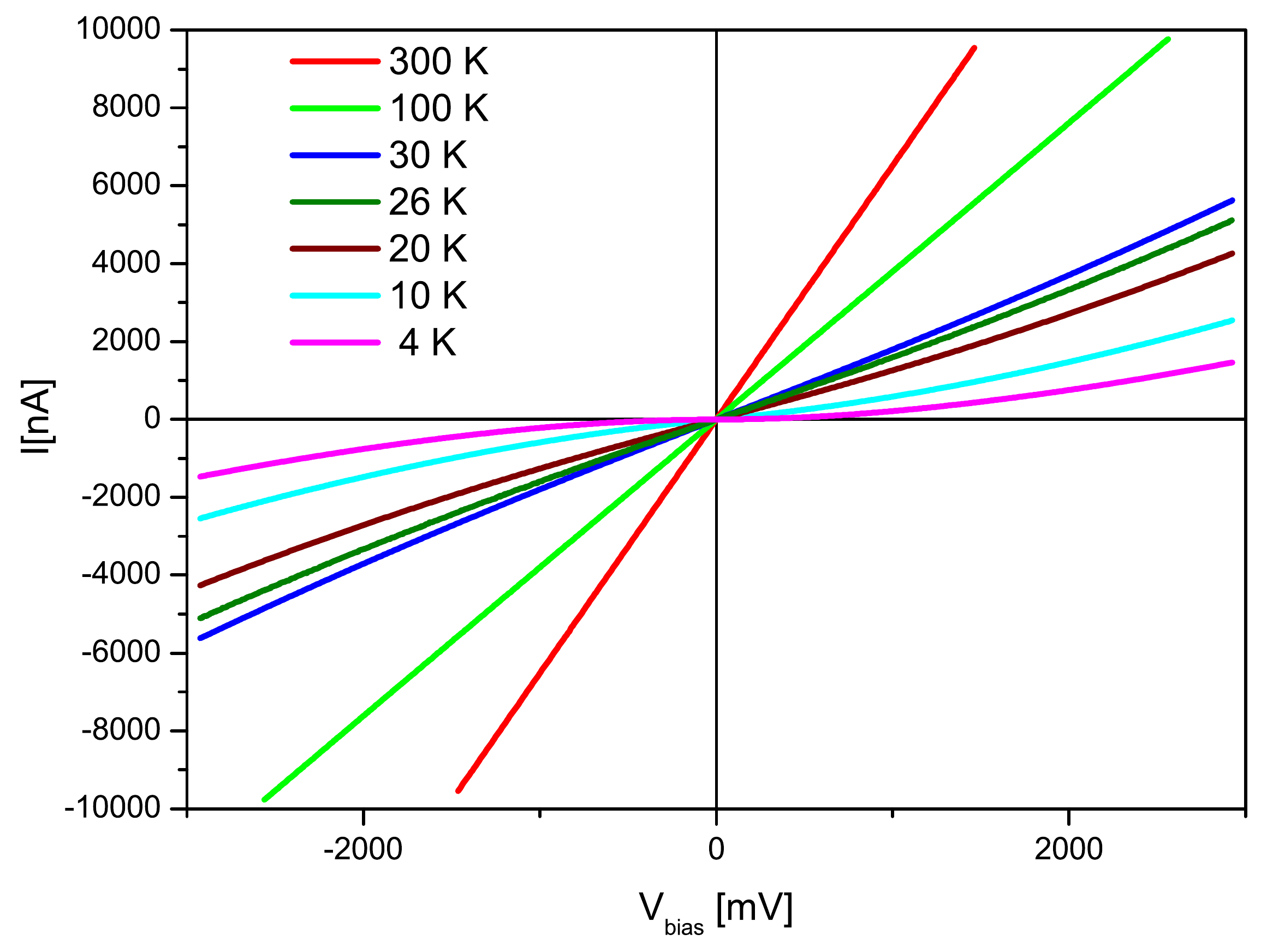}
            }
        \end{center}
\caption{ Temperature evolution of the $I-V$ curves of the Ti/TiO$_{x}$/Ti structure between 4 and 300~K.  }		
	 \label{Fig2}
			\end{figure}

%%%%%%%%%%%%%%%%%%%%%%%%%
%
			Current-voltage characteristics of selected 15~nm thick Ti/TiO$_{x}$/Ti structures in the temperature range between 4~K and 300~K were measured.
			Typical $I-V$ curves obtained are shown in Fig.~\ref{Fig2}.
			As can be seen, the $I-V$ curves are linear at temperatures from 300~K down to 30~K, whereas slopes of the
curves are strongly  temperature dependent.
			At lower temperatures, the $I-V$ curves show slight deviation from linear behaviour, indicating that tunneling through the TiO$_{x}$ barrier has only 
negligible effect on the flowing current, and dominating electrical transport is due to electrical conduction in TiO$_{x}$ material.

 %%%%%%%%%%%%%%%%%%%%%%%%%%
%%%%%%%%%%%%%%%%%%%%%%%%%
			
\begin{figure}[!b]%
				\begin{center}
											\resizebox{1\columnwidth}{!}{%
  			\includegraphics{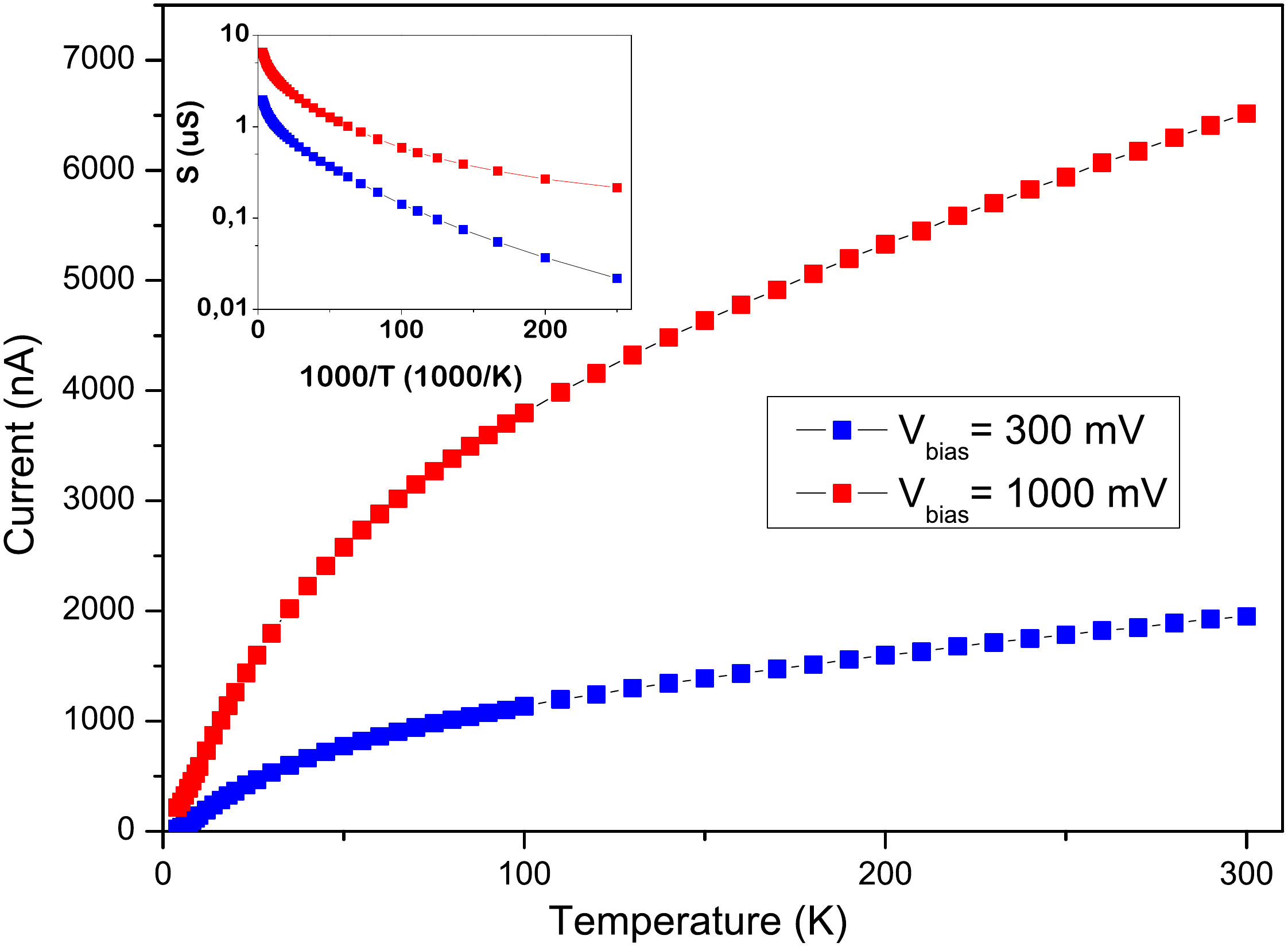}
            }
        \end{center}
\caption{ The temperature dependence of the current of the Ti/TiO$_{x}$/Ti structure between 4 and 300~K at the applied bias of 0.3 and 1~V. The inset shows corresponding conductance plotted on a logaritmic scale vs $1000/T$.
	 }		
	 \label{Fig3}
			\end{figure}

		Temperature  dependencies of the current at the applied bias voltage of 0.3~V and 1~V between 4~K  and 300~K  for the tested 
		%\linebreak 
		 Ti/TiO$_{x}$/Ti structure are shown in Fig.~\ref{Fig3}.
	The obtained curves reveal semiconducting-like temperature activated behaviour, while the current at applied constant bias voltage, $V_{bias}$, increases almost two orders of magnitude at temperature increase from 4~K to 300~K.
		As indicated by the inset that shows the conductance, $S$, for the same  values of bias voltage plotted on a logarithmic scale vs 1000/T, there is a deviation from ohmicity at low temperatures, and the temperature dependence of the conductivity, $\sigma$, can not be explained by a simple scenario of temperature activated conductivity, $\sigma \propto \exp(-E_{a}/kT)$; here $ E_{a}$ is activation energy, $k$ is the Boltzmann constant, and $T$ is temperature.

%indicates In the inset, conductance, $S$, for the same bias voltage values is plotted 
%   on a log-scale vs 1000/T, indicating deviation from ohmicity at low temperatures.
%   
%  					that temperature dependence of the conductivity, $\sigma$, can not be explained considering temperature activation (Arrhenius) law, $\sigma \propto \exp(-\Delta/kT)$ (where $\Delta$ is the activation energy, $k$ is the Boltzmann constant, and $T$ is temperature), thus analysis of the conductance data requires more complex approach.
%    
			
 %%%%%%%%%%%%%%%%%%%%%%%%%%
%%%%%%%%%%%%%%%%%%%%%%%%%
  
			Due to observed semiconducting-like behaviour, and expected high structural disorder of the investigated devices, it can be reasonably supposed that electrical conductivity of such structures at low temperatures can be adequately described in terms of variable-range hopping (VRH) between localized states in the vicinity of the Fermi energy, E$_{F}$.
			Thus,
		
\begin{equation}
\sigma = \sigma_0.\exp[-(T_0/T)^{x}],
	\label{sigma1}
 \end{equation}				 
				where the exponent $x = (m + 1 )/(m + 4)$, $T_0$ is a characteristic temperature,
		and $\sigma_0$ is a constant.
		 In the special case when density of localized states near the Fermi level is constant or depends only weakly on energy ($m = 0$; $x = 1/4$), the temperature dependence of the conductivity follows Mott's law  \cite{Mott1968,Shklovskij84}
		
\begin{equation}
\sigma = \sigma_0.\exp[-(T_0/T)^{1/4}].
	\label{sigma2}
 \end{equation}		
 %%%%%%%%%%%%%%%%%%%%%%%%%%%%%%%%% 
 On the other hand, when a parabolic 'Coulomb gap' ($m = 2$; $x = 1/2$) is created
as a consequence of Coulomb interaction, the conductivity can be described by Shklovskii-Efros law \cite{Mott1968,Shklovskij84,Ef-Sch75}.
		\begin{equation}
\sigma = \sigma_0.\exp[-(T_0/T)^{1/2}].
	\label{sigma3}
 \end{equation}
 These two specific types of VRH mechanisms are usually observed in specific temperature regions. 
 Mott's type of VRH quite often takes place at higher temperatures, and Efros-Shklovskii VRH at low temperatures, while a crossover between these two VRH mechanisms is observed   with decreasing temperature \cite{Zhang2011,Rosenbaum1997,Ghosh1998}. 

	  Performed numerical analysis of the experimental data indicates that electrical conductivity of the studied 
	   \linebreak Ti/TiO$_{x}$/Ti devices at lowest temperatures can not be satisfactorily described by VRH conductivity as expressed in general form by Eq.~(\ref{sigma1}). 
	 However, it should be taken into account that a "weak"  tip-induced oxidation process can be responsible for relatively greater fraction of TiO and other lower oxides of titanium in the formed TiO$_{x}$ region, most likely in vicinity of the substrate/TiO$_{x}$ interface. 
	 Such as TiO$_{x}$ oxides for $0.7 <  x < 1.25 $  are believed to be metallic \cite{Morin1959,Rao1974}, 
and considering that concentration of this metallic phase is sufficient to form a conductive percolative path across the TiO$_{x}$ arrea,  one can expect presence of a metallic channel in the conductance. 
	  In such case, electrical conductance of TiO$_{x}$ region can be adequately described using two-channel formula given by

	\begin{equation}
S = S_0 + A.\exp[-(T_0/T)^{x}],
	\label{S4}
 \end{equation}	
	where $S_0$ is a metallic term of conductance that is considered to be temperature independent, and second term is due to VRH conductance, in accordance with Eq.~(\ref{sigma1}).

	 %%%%%%%%%%%%%%%%%%%%%%%%%%
	\begin{figure}[!tb]%
				\begin{center}
											\resizebox{1\columnwidth}{!}{%
  			\includegraphics{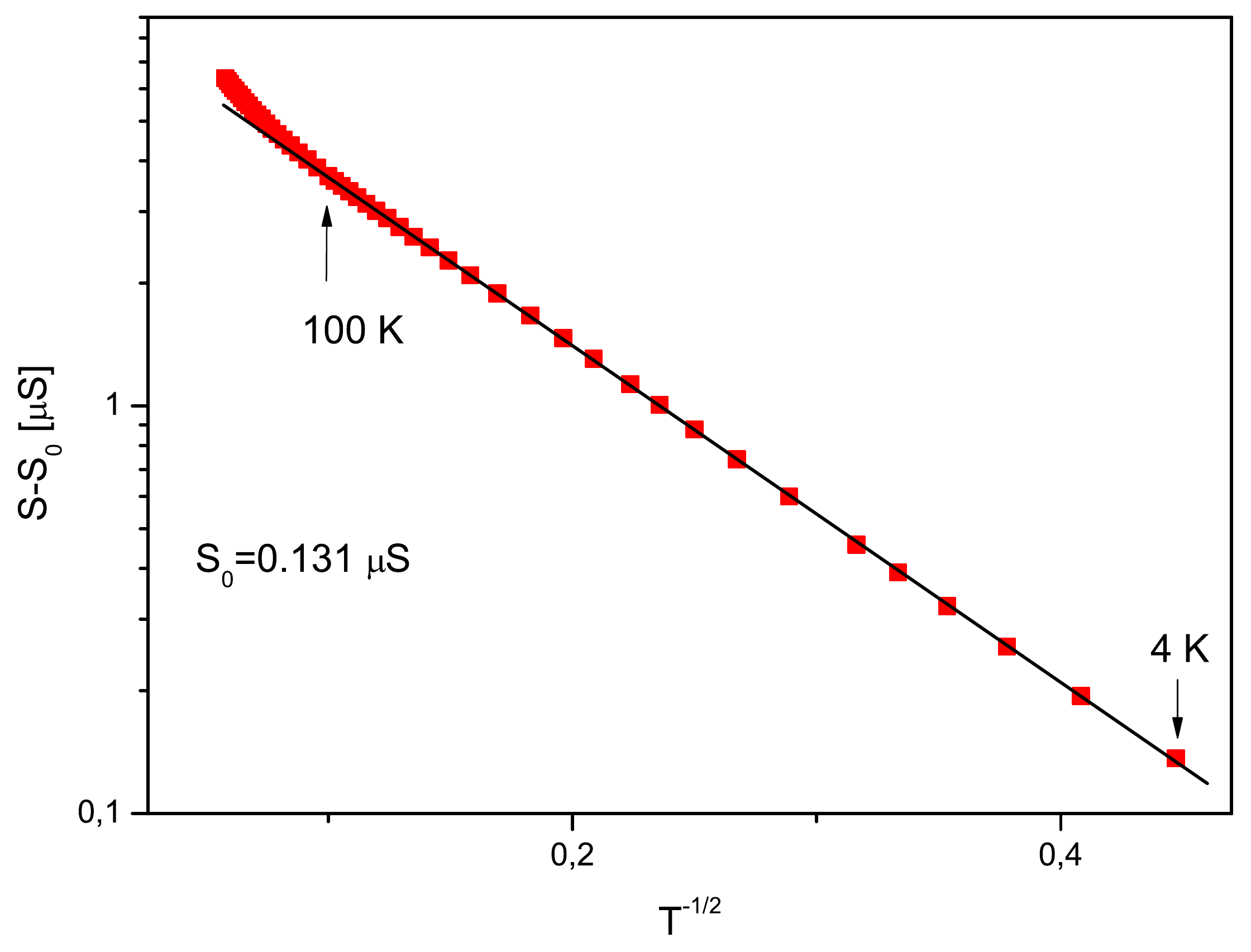}
            }
        \end{center}
\caption{ Conduction $S-S_0$ of the Ti/TiO$_{x}$/Ti structure at the applied bias of 1~V 
plotted on a log-scale {\em vs} $T^{-1/2}$
in the temperature interval  $4-300$~K. The line is obtained from linear fitting in the 
temperature range $4-100$~K. }		
	 \label{Fig4}
			\end{figure}
%%%%%%%%%%%%%%%%%%%%%%%%%
	
\begin{figure}[!h]%
				\begin{center}
											\resizebox{1\columnwidth}{!}{%
  			\includegraphics{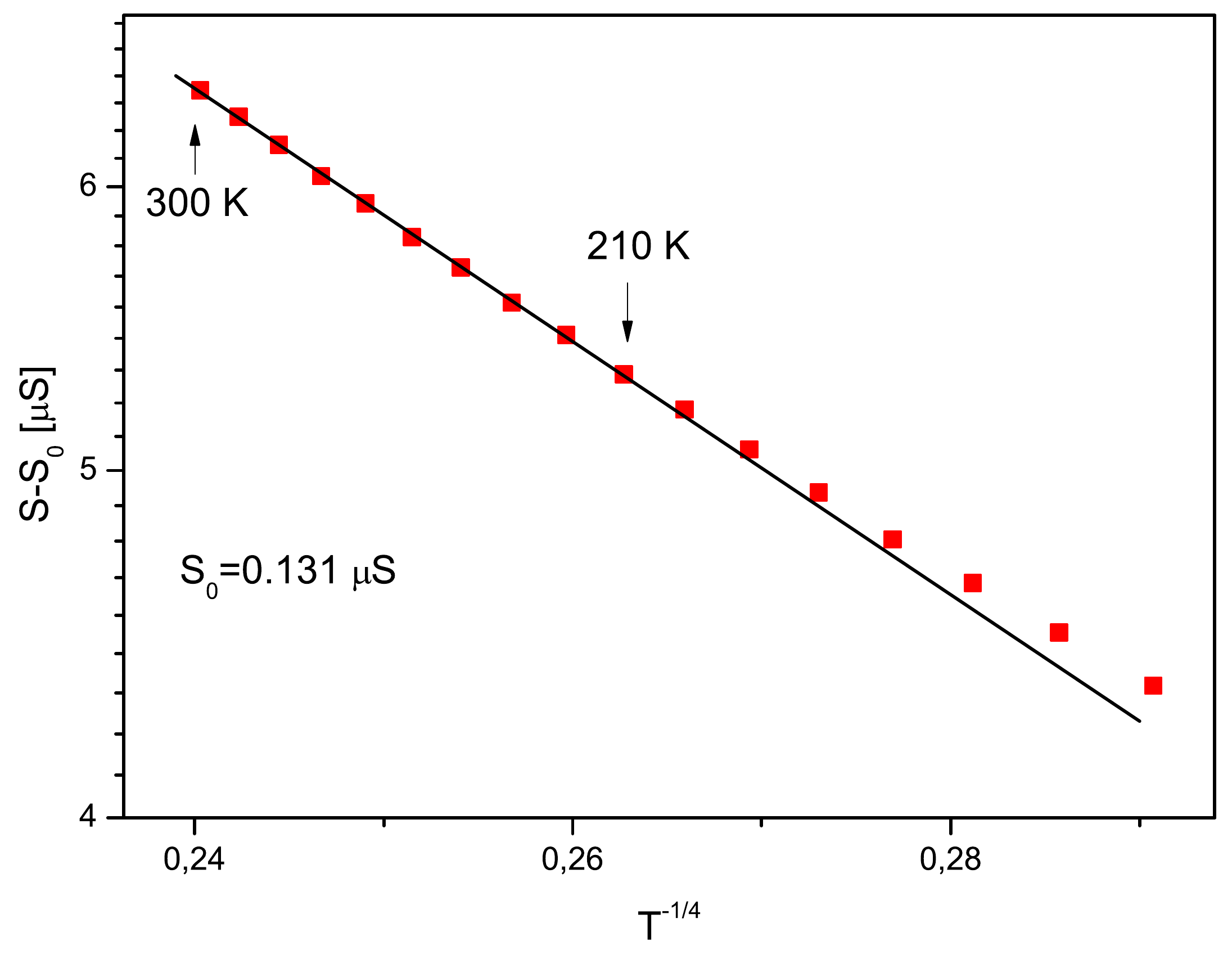}
            }
        \end{center}
\caption{Conduction $S-S_0$ of the Ti/TiO$_{x}$/Ti structure at the applied bias of 1~V 
plotted on a log-scale {\em vs} $T^{-1/4}$
in the temperature interval  $140-300$~K. The line is obtained from linear fitting in the 
temperature range $210-300$~K.}		
	 \label{Fig5}
			\end{figure}

	Indeed, as can be seen in Fig.~\ref{Fig4}, electrical conduction of the tested Ti/TiO$_{x}$/Ti structure below 100~K can be adequately described by superposition of a metallic channel and Efros-Shklowskii law in the form

		\begin{equation}
S = S_0 + S_{ES}.\exp[-(T_{ES}/T)^{1/2}],
	\label{S5}
 \end{equation}	
	where $S_0 =  0.131~\mu$S, $S_{ES} = 9.41~\mu$S, and  $T_{ES} = 90.4$~K, while at temperatures above 100~K a deviation from this behavior is observed and the curve given by Eq.~(\ref{S5}) underestimates the experimental data. 
	On the other hand, as can be seen in Fig.~\ref{Fig5}, the experimental data above 210~K can be described by superposition of a metallic channel and Mott's law in the form 
	
		\begin{equation}
S = S_0 + S_M.\exp[-(T_M/T)^{1/4}],
	\label{S6}
 \end{equation}	
 where $S_0 = $ 0.131~$\mu$S, $S_M = 45.0~\mu$S, and  $T_M =$ 4380~K.
  Note that parameter S$_0$ in equations (5) and (6) has the same value.
   This reveals a crossover from Mott VRH law at higher temperatures to Efros-Shklovskii VRH law at low temperatures.

 	 %%%%%%%%%%%%%%%%%%%%%%%%%%
%%%%%%%%%%%%%%%%%%%%%%%%%
	
\begin{figure}[!tb]%
				\begin{center}
											\resizebox{1\columnwidth}{!}{%
  			\includegraphics{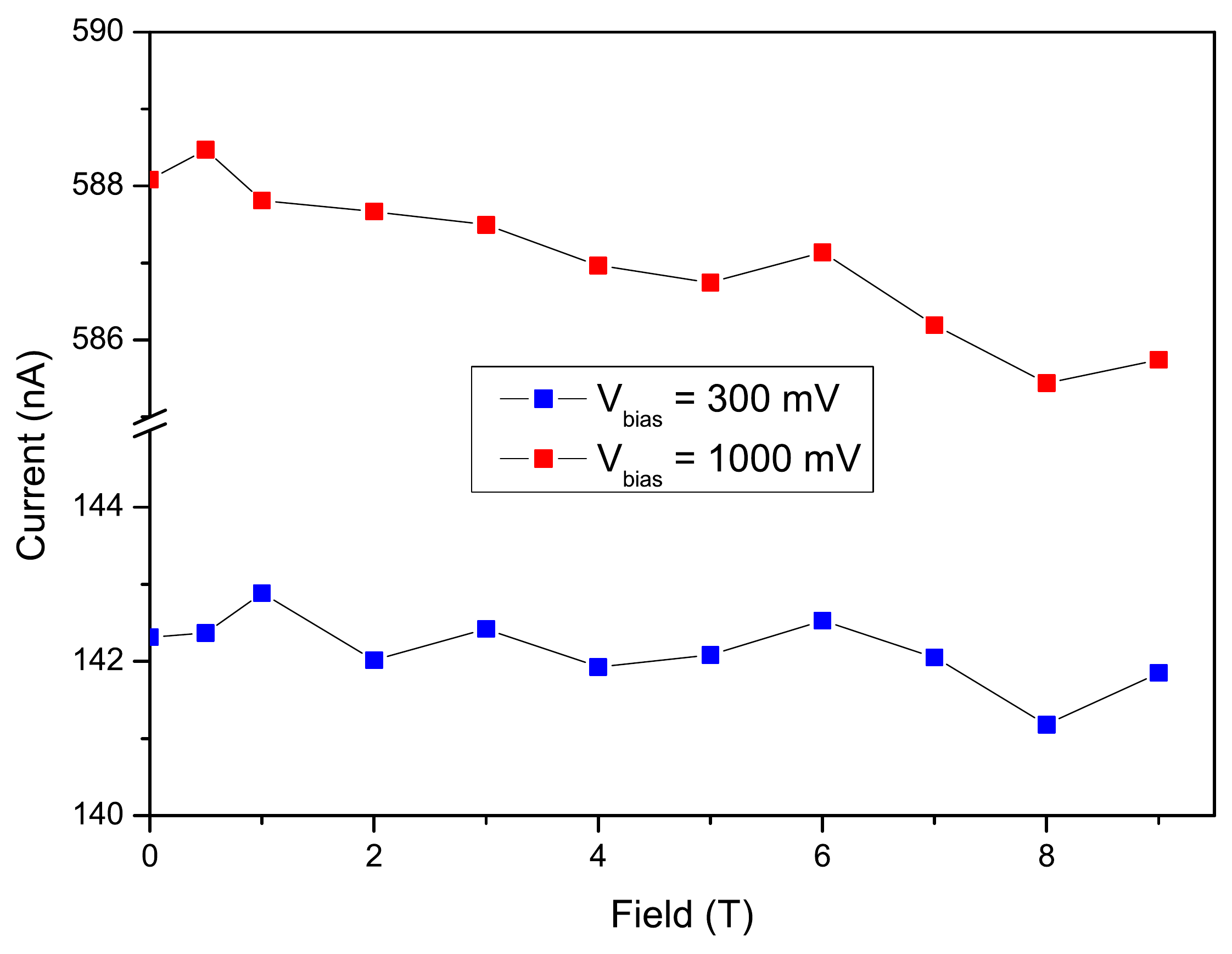}
            }
        \end{center}
\caption{Variation of the current of Ti/TiO$_{x}$/Ti structure due to applied magnetic field at the temperature of 10~K. }		
	 \label{Fig6}
			\end{figure}

 		Studying variation of the current due to applied magnetic field have shown an interesting, practical, property of the tested  structures:   
  	The Ti/TiO$_{x}$/Ti devices reveal only very small conductance changes due to applied magnetic field up to 9~T over the whole temperature region of $10 -300$~K. 
 	For example, 	as can be seen in Fig.~\ref{Fig6}, variation of the conductance in magnetic field of 8~T at the temperature of 10~K is less than 0.5~$\%$. 
 	This corresponds to the temperature shift due to magnetoconductance less than 0.4~$\%$, what is almost the same value as that of CERNOX CX-50 thermometers (produced by LakeShore Cryotronics, USA) \cite{LakeShore} at the same conditions. 
 		Such behavior in combination with the high sensitivity of the electric characteristics to temperature changes suggests a promising usage of Ti/TiO$_{x}$/Ti structures as temperature sensors for high magnetic fields.  
 		
 	Besides using as thermometers, there are many other possible applications of such structures.
 	For example, a way of preparation, e.g as described in part \ref{Experiment}, 
allows to prepare the structures  with active volume of very small mass, and correspondingly very low heat capacity for possible usage as temperature sensors with extremely fast temperature response. 
 		This can find utilization e.g. in field of (micro)calorimeters for measuring heat capacity by non-adiabatic techniques (e.g. AC-modulation \cite{Sullivan1968,Gmelin19971}, relaxation \cite{Albert1982,WILLEKERS1991}, or dual-slope technique \cite{WILLEKERS1991,Riegel1986}), as well as in monitoring temperature and/or temperature gradients at cryogenic temperatures and high magnetic fields. 
 		Moreover, the  structures can be fabricated directly on a surface of interest, and in case of need, they can be consequently passivated, e.g. by deposition of sufficiently thick SiO$_2$ layer \cite{Matsumoto1997}. 
 		Thus, it seems that Ti/TiO$_{x}$/Ti structures could be routinely fabricated and used in many applications, especially those at cryogenic temperatures and high magnetic fields.

 	 %%%%%%%%%%%%%%%%%%%%%%%%%%
%%%%%%%%%%%%%%%%%%%%%%%%%

 \section*{Summary}
In summary, 
we prepared Ti/TiO$_{x}$/Ti test structures by tip-induced LAO of titanium thin films using AFM.
The structures exhibit almost linear $I-V$ curves at temperatures from 300~K down to 30~K, and slight 
 deviation from linear behaviour is observed at the lowest temperatures.
	The electrical conduction of the structures can be reasonably described considering a metallic-type channel in parallel with a VRH channel.
		While Mott's formula of VRH is relevant  at temperatures above 210~K, Efros-Shklovskii law applies below 100~K. 
	Origin of the metallic-type conduction is associated with presence of metallic, lower titanium oxides (e.g. TiO) in the oxidized TiO$_x$ region.
	The investigated structures have revealed only small conductance changes in magnetic field, 
whith temperature shift due to magnetoconductance comparable to that of commercial	low-magnetoresistance thermometers. 
 		The observed electrical properties and extremely low mass of active arrea of the structures predetermines them 
for construction of temperature sensors with very low heat capacity for detection of temperature variations at cryogenic temperatures and high magnetic fields.
Besides this, perspective applications of Ti/TiO$_{x}$/Ti planar structures include e.g. construction of (micro)calorimeters for  
heat capacity measurements by non-adiabatic techniques or monitoring of temperature and/or temperature gradients of small objects at low temperatures.

%%%%%%%%%%%%%%%%%%%%%%%%%%%%%%%%%%%%%%%%%%%%%%%%%%%%%%%%%%%%%%%%%%%%%%%%
 \section*{Acknowledgments}
This work was supported by the VEGA project 2-0184-13, by ERDF EU (European Union European regional development fund) grant under Contract No. ITMS 26220120005 
.

%\section*{References}
%\bibliographystyle{phjcp}   %{amsplain}  %{phjcp}  {epj}

%\bibliography{referencesTiOx}
%\bibliographystyle{unsrt}  %{unsrt}

%% 
%% %\bibliography{references-Memr}

% 

%%%%%%%%%%%%%%%%%%%%%%%%%%%%%%%%%%%%%%%%%%%%%%%%%%%%%%%%%

%\bibitem{Pershin_Adv_Phys_2011}
%{\sc Y.~V. Pershin} and {\sc M.~D. Ventra},
%\newblock {\em Advances in Physics} {\bf 60}, 145 (2011).
%
%\bibitem{Mott1968}
%N.~F. Mott,
%J. Non-Cryst. Solids, \textbf{1,} 1 (1968).
%
%\bibitem{Shklovskij84}
%B.~I. Shklovskii  and A.~L. Efros  in
%  \emph{Electronic Properties of Doped Semiconductors},
%  (Springer Series in Solid State Sciences, 1984).
%
%\end{thebibliography}

\end{document}